\documentclass{article}
\topskip=\baselineskip
\let\e=\emph

\let\ct=\cite

\let\bv=\mathbf

\let\q=\widehat

\let\h=\hbar
\let\rta=\rightarrow

\let\ty=\textstyle

\newcommand{\m}{\mbox}

\newcommand{\hf}{\ensuremath{{\scriptstyle\frac{1}{2}}}}

\newcommand{\be}{\begin{equation}}
\newcommand{\ee}{\end{equation}}

\newcommand{\ba}{\begin{array}}
\newcommand{\ea}{\end{array}}
\newcommand{\bea}{\begin{eqnarray}}
\newcommand{\eea}{\end{eqnarray}}
\newcommand{\beas}{\begin{eqnarray*}}
\newcommand{\eeas}{\end{eqnarray*}}

\newcommand{\vq}{\bv{q}}

\newcommand{\vp}{\bv{p}}

\title{Orthodox quantization of Einstein's gravity: \\
       might its unrenormalizability be technically \\
           fathomable and physically innocuous?}
\author{Steven Kenneth Kauffmann \\
        American Physical Society Senior Life Member}
\date{43 Bedok Road \\
      \# 01-11 \\
      Country Park Condominium \\
      Singapore 469564 \\
      Tel \& FAX: +65 6243 6334 \\
      Handphone: +65 9370 6583 \\
      \m{} \\
      and \\
      \m{} \\
      Unit 802, Reflection on the Sea \\
      120 Marine Parade \\
      Coolangatta QLD 4225 \\
      Australia \\
      Tel/FAX: +61 7 5536 7235 \\
      Mobile:  +61 4 0567 9058 \\
      \m{} \\
      Email: SKKauffmann@gmail.com}
\begin{document}
\maketitle
\begin{abstract}
Many physical constants related to quantized gravity, e.g.,
the Planck length, mass, curvature, stress-energy, etc., are
nonanalytic in $G$ at $G=0$, and thus have expansions in powers
of $G$ whose terms are progressively more divergent with increasing
order.  Since the gravity field's classical action is inversely
proportional to $G$, the path integral for gravity-field quantum
transition amplitudes shows that these depend on $G$ only through
the product $\h G$, and are nonanalytic in $G$ at $G=0$ for the
same reason that all quantum transition amplitudes are nonanalytic
in $\h$ at $\h = 0$, namely their standard oscillatory essential
singularity at the classical `limit'.  Thus perturbation expansions
in powers of $G$ of gravity-field transition amplitudes are also
progressively more divergent with increasing order, and hence
unrenormalizable.  While their perturbative treatment is impossible,
the exceedingly small value of $\h G$ makes the semiclassical
treatment of these amplitudes extraordinarily accurate, indeed to
such an extent that purely classical treatment of the gravity field
ought to always be entirely adequate.  It should therefore be fruitful
to couple classical gravity to other fields which actually need to be
quantized: those fields' ubiquitous, annoying ultraviolet divergences
would thereupon undergo drastic self-gravitational red shift, and thus
be cut off.
\end{abstract}

\subsection*{Introduction}

Gravity occupies a profoundly fundamental place in physical theory by
virtue of its source being the energy-momentum tensor, whose presence
accompanies \e{any} physical phenomenon whatsoever.  This
\e{universality} implies that the source of the gravitational field
even includes a contribution from \e{itself}, which is the origin of
its \e{nonlinearity}.  Gravity's predominantly \e{negative} long-range
field potential energy response to its sources can enable it to
overwhelm and suppress those source components that happen to be
sufficiently concentrated and strong, even plunging them into ``black
holes'' in extreme cases, but its \e{fundamental} coupling strength is
\e{smaller} than that of the other known forces to a mind-boggling
degree: the mutual gravitational attraction between two even
neutron-rich nuclei (e.g., tritons) is still weaker than their mutual
Coulomb repulsion by a staggering factor of over $10^{35}$.

As Einstein's gravitational theory and the quantum theory of
dynamics are both crowning achievements of twentieth-century
theoretical physics, it is entirely natural to try to combine them.
The multitude of similarities between Einstein's gravity theory and
Maxwell's electrodynamics, together with the qualifiedly successful
(at least after `renormalization') quantized treatment of the latter
in the approximation context of a perturbation expansion in powers of
the coupling strength, strongly suggests handling the quantization of
gravity analogously, \e{especially} in view of the fact that its coup%
ling strength is even \e{vastly weaker} than that of electrodynamics.
Unfortunately, however, as has been realized for well over half a
century~\ct{Wn}, the perturbation expansions in powers of the grav%
itational coupling strength $G$ of transition amplitudes of the canon%
ically quantized gravitational field yield infinities whose severity
increases progressively with order, which precludes `renormalization'.
Since `renormalization' consists of a certain class of prescriptions
that are imposed \e{atop} a perturbatively treated quantum field
theory for the express purpose of \e{shunting aside} unwelcome
infinities that it has produced (``sweeping them under the rug'',
in Feynman's blunt phrase), but \e{fails} to point to \e{any physical
mechanism} which disallows their existence, its \e{inapplicability}
to canonically quantized gravity is obviously \e{not of itself}
adequate reason to conclude that that theory must be discarded---%
\e{especially} in light of the \e{robustness} of the twin pillars,
namely Einstein's gravity and quantum dynamics, upon which it rests.

That notwithstanding, it was promulgated as ``pragmatically
motivated'' dogma in the late 1960's that \e{all} unrenormalizable
quantum field theories were henceforth to be regarded as being
``beyond the pale'' on the grounds that if the \e{perturbation
expansion} for a quantum field theory was \e{unviable}, it simply was
\e{not conceivable} that any \e{practicable alternate means}
of satisfactorily extracting its results could exist or be developed.
This ``pragmatic'', but obviously \e{not physically based}, dogma,
which \e{also} directly flouts Einstein's admonition not to cravenly
limit research efforts to ``the thinnest part of the board'', then
drove a single-minded effort to create renormalizable theories at
essentially any cost: Occam's razor and its companion guidelines of
conservatism, continuity, and known empirical support for undertaking
modification of physical theory were upended as fields were abruptly
swapped for strings, the four dimensions of space-time were simply
increased to ``whatever it takes'', and fermionic Noether currents
which anticommute rather than causally commute at spacelike
separations were ascribed with physical existence---all done \e{not}
to accommodate existing \e{physical knowledge} but simply in
brainstorming support of a headlong tunnel-vision effort to create
theoretical structures whose perturbative infinities are ``under
control'' in a particular preconceived sense.

It was not \e{inevitable}, however, that the catastrophic failure of
the perturbation expansion in powers of $G$ for canonically quantized
gravity need have so abjectly contributed to the above-described
departures from the guidelines which have well served theoretical physics
research for centuries.  The keen sense of frustration which arose from
that failure \e{ought} to have been tempered by the realization that
theories can be qualitatively probed with simple, if somewhat blunt tools
that at least have the virtue of being entirely \e{nonperturbative}.  One
such rough tool is the time-honored one of dimensional analysis---this seems
almost made to order for the canonically quantized gravitational field
with its highly suggestive ``Planck trio'' of applicable constants, namely
$G$, $c$, and $\h$.  It is well-known that from these universal constants
Planck entities having the dimensions of mass, length, and time are readily
constructed, namely the Planck mass $\sqrt{\h c/G}$, the Planck length
$\sqrt{\h G/c^3}$, and the Planck time $\sqrt{\h G/c^5}$, which, in turn,
are basic building blocks of further Planck entities that have any dimen%
sion one might wish to nominate (with the exception of dimensionlessness).
Such entities are of particular interest to us for extremely \e{small}
values of $G$, which accord with both $G$'s actual \e{physical} value
\e{and} with the point $G=0$ about which the disastrous perturbation
expansion of canonically quantized gravity is made.  We are immediately
struck by the fact that \e{all three} of the basic Planck entities given
above, notwithstanding that they are obviously \e{perfectly well-defined}
(indeed elementary), nevertheless have perturbation expansions in powers
of $G$ whose terms become progressively more severely divergent with
increasing order---which is precisely the \e{same} ``perturbatively
catastrophic'' behavior that has for so many decades been regarded as the
death knell of gravity's straightforward canonical quantization!  It now
becomes apparent that canonically quantized gravity \e{may} have been
\e{prematurely} written off all those decades ago on entirely inadequate
grounds---any otherwise well-defined quantity that, like the above three
basic Planck entities, is \e{nonanalytic} in $G$ at the point $G=0$, will
normally have just such a ``catastrophic'' perturbation expansion in powers
of $G$.  There are, indeed, many \e{other} physically interesting (and
perfectly well-defined) Planck entities that \e{also} fall precisely into
this category, including, inter alia, the Planck curvature $c^3/(\hbar G)$,
Planck energy density $c^7/(\hbar G^2)$, Planck acceleration
$\sqrt{c^7/(\hbar G)}$, and Planck wave number $\sqrt{c^3/(\hbar G)}$.
We now need to inquire into why $G=0$ is a point of nonanalyticity of so
many physical entities which flow from canonically quantized gravity
(conceivably including, it now seems not implausible, its transition
amplitudes).

\subsection*{Small $G$ and the classical limit}

We see that as $G$ tends toward zero, the Planck wave number just mentioned
increases without bound, which at least \e{suggests} that $G \rta 0$
drives the canonical quantization of gravity toward its \e{classical limit}.
Furthermore, in line with what would be expected of a wave number marker for
the classical limit, the Planck wave number \e{also} increases without bound
as $\h$ tends toward zero---in fact, the Planck wave number depends on $G$
through the \e{product} $(\h G)$.  It is, of course, well-known that
quantum theories behave in an extremely nonsmooth asymptotic fashion as
they are driven to their classical limit (e.g., when $\h \rta 0$), so we
now glimpse fragments of an argument as to why many physical entities
which flow from canonically quantized gravity might be expected to be
nonanalytic in both $\h$ at $\h =0$ and in $G$ at $G=0$.  In
order to present that argument in a clear, systematic way for this theory's
transition amplitudes in particular, we shall first review the reasons why
quantum transition amplitudes \e{in general} are normally nonanalytic in
$\h$ at $\h =0$.  We shall also briefly discuss a path-integral-based
stationary phase asymptotic semiclassical expansion approach to quantum
transition amplitudes which is valid as $\h \rta 0$.

General quantum transition amplitudes
such as $\langle\psi_f|\exp (-i\q{H}%
(t_2-t_1)/\hbar)|\psi_i\rangle$ can be rewritten in
terms of the eigenspectrum of $\q{H}$ as $\sum_E%
\langle\psi_f|E\rangle\exp (-iE(t_2-t_1)/\hbar)%
\langle E|\psi_i\rangle$, and therefore will
obviously almost \e{always} be \e{nonanalytic} in
$\hbar$ at $\hbar=0$.
A systematic treatment of such transition amplitudes
as $\hbar$ approaches their point of nonanalyticity
at $\hbar=0$ can be developed from their \e{path
integral expression},
\begin{eqnarray*}
\lefteqn{\langle\psi_f|\exp (-i\q{H}(t_2-t_1)/\hbar)|\psi_i\rangle= } \\
  & & \int d^n\vq_2\:\langle\psi_f|\vq_2\rangle
      \int d^n\vq_1\:\langle\vq_1|\psi_i\rangle\times \\
  & & \int\mathcal{D} {\textstyle\{\, (\vq (t),\vp (t))\;\, |\;\: t\,\epsilon\,
      [t_1,t_2],\vq (t_1)=\vq_1,\vq (t_2)=\vq_2\} }\times \\
  & & \mbox{\hspace*{1.3em}}{\textstyle\exp(i\int_{t_1}^{t_2}dt\,
      (\dot\vq (t)\cdot\vp (t)-H(\vq (t),\vp (t)))/\hbar)},
\end{eqnarray*}
which again makes their \e{nonanalyticity} in $\hbar$
at $\hbar=0$ manifest.  As $\hbar\rightarrow 0$, how%
ever, the pure phase integrand of the path integration
oscillates increasingly rapidly except in an increas%
ingly small neighborhood of the path which renders it
\e{stationary}.  This stationary path is readily seen
to satisfy $\dot\vq (t)=\nabla_{\vp (t)}H(\vq (t),\vp (t))$
and $\dot\vp (t)=-\nabla_{\vq (t)}H(\vq (t),\vp (t))$,
which are, of course, Hamilton's \e{classical} equations
of motion, subject to the end constraints imposed on the
paths integrated over, namely that $\vq (t_1)=\vq_1$ and
$\vq (t_2)=\vq_2$.  The \e{full} mathematical development
of this sort of stationary phase approximation to an
integral over a pure phase integrand which is driven
by a parameter to oscillate increasingly rapidly is
well known to produce an \e{asymptotic expansion} of
the integral in that parameter.  To be sure, this
stationary phase asymptotic expansion technology is
not normally presented in the context of \e{path}
or \e{functional} integrals, but all the needed con%
cepts and theorems, including polynomials, derivatives,
the Taylor expansion, Gaussians, and analytic integra%
tion of the products of polynomials with Gaussians, are
readily extended from the mathematics of multivariate
functions to that of functionals.  This stationary
phase asymptotic expansion technique applied to
the quantum path integral, with $\hbar\rightarrow 0$
being the driver of the increasingly rapid pure phase
integrand oscillations, provides the systematic
\e{semiclassical} expansion of \e{quantum transition
amplitudes}---a methodology of considerable promise
for strong interactions, which \e{has not yet} been tapped.
It is to be cautioned, however, that the approach may
be a formidable consumer of computational resources,
as it, in principle, requires all the classical paths
generated by \e{every possible pair} of end constraints.

For particle dynamics, as it is treated above, the classical
action functional is,
\[{\ty\int_{t_1}^{t_2}}dt\, (\dot\vq(t)\cdot\vp (t) - H(\vq (t),\vp (t))).\]
For the gravitational field, however, it turns out that
the classical action functional is \e{inversely}
proportional to $G$, i.e., it equals the constant factor
$(-c^4/(16 \pi G))$ times the curvature scalar integrated over
generally invariant space-time~\ct{Wn1}.  Therefore for the canonically
quantized gravitational field, $(16 \pi \hbar G)/c^4$ may be expected to
play a role analogous to that played by $\hbar$ \e{alone}
in quantized dynamics generally.  Thus we may \e{indeed}
expect canonically quantized gravitational field theory to be
\e{nonanalytic} in $G$ at $G=0$, as well as in $\h$ at $\h =0$,
and its perturbation expansion in powers of $G$ to be a disaster.
However, since $(16 \pi \hbar G)/c^4$ is notable for
its extreme smallness, we may \e{also} expect the stationary phase
\e{semiclassical} asymptotic expansion of canonically
quantized gravitational field theory to produce
extraordinarily accurate results indeed.  In fact, this
approach may be expected to yield \e{such} good results that
simply resorting to the purely \e{classical} gravitational field
ought to be entirely adequate.

Another way to appreciate the predominantly \e{classical}
character of the canonically quantized gravitational field
is to consider the detectability of individual gravitons.
The extreme weakness of the gravitational coupling strength
$G$ makes individual gravitons essentially undetectable
\e{unless} they have extraordinarily high energy.  Any process
capable of emitting such gravitons would almost certainly
involve extremely strong gravitational fields in the
immediate vicinity of their region of emission, fields
which would tend to gravitationally red shift those very
gravitons to lower energy.  It thus might be problematic
for gravitons energetic enough to be individually detect%
able to actually be available.  Furthermore, the total
phase space for a graviton to decay into two gravitons
that both travel in its original direction is nonvanish%
ing (albeit for three or more gravitons it does vanish).
This decay is suppressed both by the weakness of $G$ and
by a d-wave orbital angular momentum barrier, but its
rate should rise strongly with energy, thus also deplet%
ing the availability of gravitons energetic enough to be
individually detectable.  (Interestingly, two-photon de%
cay of a photon is ruled out in spin $\hf$ quantum elec%
trodynamics by Furry's theorem.) Finally, the distinctly
\e{macroscopic} magnitudes of the Planck mass (which at
nearly 22 mcg is comparable to that of a small puntuation
mark cut out of a glossy page), the Planck momentum (which
at over 23,000 g km/hr is comparable to that of a bullet),
and the Planck energy (which at over 540 kWh would supply
a household for many days) hardly suggest a significant
need to take \e{quantum} corrections to the
gravitational field into account.

It is quite interesting that in the course of pondering
the \e{quantization} of the gravitational field, one is
driven to the conclusion that this endeavor is largely
\e{unnecessary}.  Einstein even more fiercely opposed
the quantization of gravity than he opposed the quantum
theory generally---as fate would have it, quantized gravity
theory \e{itself} turns out to be disinclined to disagree with
him to any \e{significant} extent.  The dominantly \e{classical}
character of \e{universal} gravitation turns out to provide a
deep validation of the Copenhagen interpretation of quantum
theory: in principle any `pure' quantum state is necessarily
(albeit usually \e{extremely} weakly!) coupled to the
\e{universal} gravitational field, which is effectively a
``classical observer'' that must in due course bring about
the ``collapse'' of its quantum coherence.  What a pity that
Einstein never confronted the reverberating ironies implicit in
this line of thought!  (In actual practice, of course, the vastly
more strongly coupled electromagnetic field is \e{very} much more
likely to play this ``quantum coherence collapsing'' role, but
electromagnetism is in principle neither universally coupled nor
necessarily dominantly classical.)

\subsection*{Dominantly classical gravity in a quantum world}

This dominantly classical character of gravitation
sorely needs, however, for the purposes of theoretical
physics, to be appropriately conjoined with the markedly
\e{quantum} characteristics which so many other physical
phenomena, such as electromagnetism, can manifest.  A straight%
forward formal approximation technology for accomplishing
this has been proposed by Boucher and Traschen~\ct{B-T}%
, wherein a hybrid partially quantized Hermitian density
operator is taken to be merely a \e{function} of those
phase-space variables (e.g., the gravitational ones)
which are to be left as unquantized c-numbers.  The same
hybridization applies to the Hamiltonian and other dynam%
ical variables of interest.  The equation of motion of
the hybrid density then is taken to involve a natural
hybrid commutator-cum-Poisson bracket of that density
with the hybrid Hamiltonian (the factors of the Poisson
bracket part of this hybrid bracket must, of course, be
ordered so as to ensure the hybrid bracket's Hermiticity).
Albeit straightforward and natural, this approach \e{cannot}
be regarded as the realization of some manner of quantum/%
classical dynamical `subtheory', because there is no \e{guar%
antee} that its hybrid density remains \e{positive} as time
evolves, as pointed out by Boucher and Traschen~\ct{B-T}%
.  Therefore this approach definitely falls in the category
of being an \e{approximation} technology with the character%
istic property of being subject to manifest failure \e{if}
applied well \e{beyond} its appropriate scope.  It is in fact
impossible for classical degrees of freedom to \e{strictly}
maintain their quintessential \e{inherent determinism} once
they are permitted to \e{interact} with \e{quantum} degrees
of freedom that are \e{not} bound by such determinism.

In a much less methodical vein, it is instructive to try to
tease out in a rough, qualitative manner some of the
salient implications of the predominantly classical gravita%
tional field for quantum phenomena.  A driving goal of high
energy particle physics is to resolve natural phenomena at
ever smaller spatial scales.  To resolve an object of extre%
mely short length $l$, we need quanta of of momenta around
$\hbar/l$ or larger to be absorbed and then reemitted (i.e.,
scattered) by that object.  As we suppose $l$ to have ever
smaller values, we may safely assume such quanta to be ultra%
relativistic, i.e. photon-like.  Upon the quantum's absorp%
tion, the object of length $l$ will have an energy of at
least $\hbar c/l$, which will generate a dimensionless
gravitational potential of around $-\hbar G/(c^3l^2)$ at its
extremities, and this, in turn, would tend to reduce the
momentum of the \e{reemitted} quantum by the factor $(1-(\hbar %
G)/(c^3l^2))$ because of gravitational redshift.  (We have used
very crude Newtonian-like gravitational guesstimates here---%
these do not include nonstatic corrections nor take account
of the needed self-consistency iterations.)  The thrust of this
crude exercise is clearly that a short \e{enough} length will
be very difficult to resolve, as the requisitely high momentum
quantum will, after absorption into the object of this length,
tend to redshift its \e{reemitted} counterpart toward extinc%
tion.  So the very \e{means} of resolving a small enough region
has the side effect of redshifting \e{itself} (upon reemission)
toward nonexistence---the \e{necessarily} energetic probe drives
its tiny target in the direction of becoming an invisible `black
hole'.  The above expressions strongly suggest that this effect
will indeed `bite' when the target length $l$ is significantly
less than $\sqrt{\hbar G/c^3}$, the Planck length.  Furthermore,
if it is not possible to resolve length intervals significantly
smaller than the Planck length, it is rather clear that `stop%
watches' which reliably record time intervals significantly
shorter than the Planck time $\sqrt{\hbar G/c^5}$ cannot be
constructed either.  Thus we would expect space-time below
the Planck scale to be not so much a `quantum foam' as
\e{intractably opaque}.  A less crude, more detailed exposition
of this argument is to be found in Ng and van Dam~\ct{N-D}.

In many quantum field theories, such as quantum electrodynamics,
the presence of virtual particles of \e{arbitrarily large} ener%
gy can be a source of mathematical divergences known as the
`ultraviolet catastrophe'.  A virtual particle of very high
energy $E$, however, can only exist for a very short time of
order $\hbar /E$ before it must be reabsorbed.  Hence its evan%
escent presence will have been confined to a region whose
length is around $\hbar c/E$.  Again resorting to very crude
Newtonian-like gravitational guesstimation, we obtain that it
will have given rise to an average dimensionless gravitational
potential of around $-GE^2/(3\hbar c^5)$ in that region, which
roughly reduces its energy from $E$ to $E(1-GE^2/(3\hbar c^5))$.
We would therefore expect virtual particle energies $E$ to be
limited to being not greatly larger than the Planck energy
$\sqrt{\hbar c^5/G}$, or else the virtual particle tends to
disappear entirely into a black hole of its own making.
This gravitational limit to virtual particle energies in turn
yields a gratifying natural cutoff for the `ultraviolet catastrophe'
divergences---the correctness of the basic thrust of this universal
natural cutoff idea for the `ultraviolet catastrophe' divergences
has recently received some detailed support in the case of
quantized scalar fields via a model wherein each virtual scalar
particle is subjected to the gravitational fields produced by its
virtual companions~\ct{Ca}.  Crudely inserting such a Planck-scale
cutoff into the divergent electromagnetic correction to the
electron's bare mass in QED yields a result roughly comparable to
the bare mass itself.  For a charged spin 0 particle, however,
the electromagnetic mass contribution would be within about
an order of magnitude of the Planck mass, independent of the
particle's bare mass.  Perhaps not so coincidentally, the
known charged spin 0 particles are all believed to be
\e{composed} of charged spin $\hf$ particles (i.e., quarks).
In contrast, the divergent apparent corrections to the
electron's \e{charge} must be purely \e{artifacts} of the
unphysical `ultraviolet catastrophe', since even \e{virtual}
processes are \e{just} as formally constrained to \e{conserve
charge} in QED as they are to respect gauge invariance.  Gauge
noninvariant infinities in QED are recognized as unphysical
artifacts that need to be \e{subtracted out}, so charge
nonconserving infinities must be handled \e{likewise},
but unfortunately they were \e{enshrined} very early on as
infinite `charge renormalizations' in flawed analogy with
``effective charge reductions'' \e{within} polarizable media.%
\footnote{This very useful way to view the electrostatic inter%
action of two probe charges \e{within} a polarizable medium
obviously \e{falls away} once the probe charge separation be%
comes \e{much larger} than the dimensions of the polarizable
specimen being considered---as their separation$\:\rightarrow\!
\infty$, that separation squared times the electrostatic force
on at least one of the probe charges approaches the \e{same}
limit as in the \e{absence} of the polarizable specimen.
In QED an infinite `charge renormalization' artifact
\e{already} occurs for a \e{single} virtual electron-positron
pair that has a vanishingly small spacelike total four-momentum.
Because this virtual pair is ``off mass shell'' by at least
\e{two electron masses}, and therefore must annihilate \e{within
the corresponding time} $\hbar /(2m_{\mathrm{e}}c^2)$ to recreate
the virtual photon whose dissociation gave it birth, its effect%
ive ``polarizable specimen dimension'' can encompass no more
than about half an electron Compton wavelength.  The minuscule
extent of this ``virtual pair polarizable specimen''---no larger
than an electron's \e{own} quantum relativistic size---makes it
obvious that this speck \e{cannot}, even during its fleeting
existence, have effected \e{any} `charge renormalization', let
alone an \e{infinite} amount, so that phenomenon clearly \e{must}
be an ultraviolet divergence \e{artifact}.}  Once classical
gravitation and its consequent virtual particle energy cutoff
have been properly incorporated into QED such that charge conser%
vation continues to be respected, there can be no alternative but
for infinite `charge renormalizations' to thereupon \e{vanish id%
entically}, which is precisely what calculations confirm~\ct{Ka}.

\subsection*{Conclusion}

The approaches of Refs.~\ct{Ca} and \ct{Ka} to the approximate in%
corporation of classical gravitation into quantum field theory fall
short of being physically fully systematic: the approach of Ref.~%
\ct{Ca} fails to include the \e{self}-gravitational phenomenon,
crudely described above, which affects a virtual particle that is
sufficiently far off its mass shell, while the approach of Ref.~%
\ct{Ka} is very dependent on a stress-energy related technical
characteristic that is peculiar to photon virtual-dissociation
Feynman diagrams.

The hybrid density operator approach of Boucher and Traschen~%
\ct{B-T} would seem to hold out the best hope for systematically
taking into account approximate classical gravitational effects
that can be spontaneously provoked by quantized fields.  However,
since the Feynman-diagram approach to quantized fields was not de%
veloped in the formal context of the density operator, the formal
calculational basis of quantum field theory in this context must
be worked out.  In partial compensation for this onerous task it
may well not be necessary that every contribution to transition
amplitudes be expressed in manifestly Lorentz invariant form in
light of the fact that classical gravity seems to be very effective
in eliminating ultraviolet divergences wherever and however they
occur~\ct{Ca, Ka}.  But in addition to this welcome removal of ul%
traviolet divergences, taking systematic account of classical self%
-gravitational effects will as well generate a very unwelcome myri%
ad of other gravitational corrections that are completely negligi%
ble---an at least semiautomatic way to avoid the labor of calculat%
ing these without distorting gravity's effect on the ultraviolet
divergences themselves will almost certainly need to be developed.
Thus some very difficult tasks will have to be addressed before
classical gravity's role in eliminating ultraviolet divergences
from quantum field theories can be elucidated in physically syste%
matic fashion.

\end{document}